\documentclass[a4paper,11pt]{article}

\usepackage{hyperref}%
\hypersetup{%
  colorlinks=true,%
  linkcolor=blue,%
  citecolor=blue,%
  urlcolor=black,%
  bookmarksnumbered=true,%
  bookmarkstype=toc,%
  bookmarksopen=false,%
  pdftitle={Discrete Scale-Invariant Boson-Fermion Duality in One Dimension},%
  pdfkeywords={boson-fermion duality; identical particles; discrete scale invariance; exact solutions},%
  pdfauthor={Satoshi Ohya}}%

\usepackage[margin=1in]{geometry}%

\usepackage[tt=false,sb]{libertine}%
\usepackage[T1]{fontenc}%
\usepackage{libertinust1math}%
\usepackage[cal=stix,scr=boondoxo,bb=boondox]{mathalpha}%
\usepackage{bm}%
\DeclareMathOperator{\e}{e}%
\DeclareMathOperator{\sgn}{sgn}%
\allowdisplaybreaks[1]%

\usepackage{enumitem}

\usepackage{epic,eepic,xcolor}%

\usepackage[font=small,hang]{caption}%
\usepackage[labelformat=simple]{subcaption}%
\usepackage[backend=biber,style=phys,biblabel=brackets,block=ragged,eprint=true,url=true]{biblatex}%
\title{\Large\bfseries Discrete Scale-Invariant Boson-Fermion Duality in One Dimension}%
\author{\normalsize Satoshi Ohya\\[1em]
  \small\itshape Institute of Quantum Science, Nihon University,\\
  \small\itshape Kanda-Surugadai 1-8-14, Chiyoda, Tokyo 101-8308, Japan\\[1ex]
  \small\ttfamily ohya.satoshi@nihon-u.ac.jp}%
\date{\small(Dated: \today)}%

\begin{document}
\maketitle%
\flushbottom%

\begin{abstract}
  We introduce models of one-dimensional $n(\geq3)$-body problems that
  undergo phase transition from a continuous scale-invariant phase to
  a discrete scale-invariant phase. In this paper, we focus on
  identical spinless particles that interact only through two-body
  contacts. Without assuming any particular cluster-decomposition
  property, we first classify all possible scale-invariant two-body
  contact interactions that respect unitarity, permutation invariance,
  and translation invariance in one dimension. We then present a
  criterion for the breakdown of continuous scale invariance to
  discrete scale invariance. Under the assumption that the criterion
  is met, we solve the many-body Schr\"{o}dinger equation exactly; we
  obtain the exact $n$-body bound-state spectrum as well as the exact
  $n$-body S-matrix elements for arbitrary $n\geq3$, all of which
  enjoy discrete scale invariance or log-periodicity. Thanks to the
  boson-fermion duality, these results can be applied equally well to
  both bosons and fermions. Finally, we demonstrate how the criterion
  is met in the case of $n=3$; we determine the exact phase diagram
  for the scale-invariance breaking in the three-body problem of
  identical bosons and fermions. The zero-temperature transition from
  the unbroken phase to the broken phase is the
  Berezinskii-Kosterlitz-Thouless-like transition discussed in the
  literature.
\end{abstract}

\newpage
\section{Introduction}
\label{section:1}
Discrete scale invariance, or scale invariance with respect to one
particular scale, has attracted considerable attention in many
scientific disciplines \cite{Sornette:1997pb,Ovdat:2019ywg} because of
its unique yet universal predictions. For example, in quantum
scattering theory, discrete scale invariance manifests itself in
log-periodic oscillations \cite{Saleur:1996} of S-matrices and in
geometric scaling of bound-state energies. Let us first take a brief
look at these ideas by using a toy example.

Consider a $1\times1$ S-matrix $S(E)$ in a specific channel, where $E$
stands for energy. The most general scaling law that respects the
unitarity $|S(E)|=1$ would have the following form:
\begin{align}
  S(\e^{t}E)=S(E),\label{eq:1}
\end{align}
where $t$ is a real parameter. If this holds for any continuous
$t\in\mathbb{R}$, the general solution to Eq.~\eqref{eq:1} must be
independent of the modulus of $E$; that is, the S-matrix is a constant
in continuous scale-invariant theory. On the other hand, if
Eq.~\eqref{eq:1} holds only for some discrete
$t\in t_{\ast}\mathbb{Z}=\{0,\pm t_{\ast},\pm2t_{\ast},\cdots\}$,
where $t_{\ast}$ defines one particular scale, the general solution
becomes $S(E)=f(\log E)$, where $f$ is a periodic function with period
$t_{\ast}$; that is, if continuous scale invariance is broken to
discrete scale invariance, the S-matrix exhibits periodic oscillations
as a function of $\log E$.

In addition to this log-periodicity, discrete scale invariance also
leads to a striking consequence in bound-state problems. Suppose that
the S-matrix has a bound-state pole along the negative $E$-axis; that
is, $S(E)\to\frac{N_{\ast}}{E+E_{\ast}}$ as $E\to-E_{\ast}$, where
$E_{\ast}>0$ and $N_{\ast}$ are some constants. Then, the scaling law
$S(E)=S(\e^{nt_{\ast}}E)$ implies that there in fact exist infinitely
many poles of the form
\begin{align}
  S(E)\to\frac{N_{\ast}\e^{-nt_{\ast}}}{E+E_{\ast}\e^{-nt_{\ast}}},\quad n\in\mathbb{Z}.\label{eq:2}
\end{align}
Hence, in bound-state problems, discrete scale invariance manifests
itself as the onset of infinitely many bound states with the energies
$E_{n}=-E_{\ast}\e^{-nt_{\ast}}$, which satisfy the geometric scaling
$E_{n+1}=\e^{-t_{\ast}}E_{n}$. Notice that the residues of the
S-matrix \eqref{eq:2}, which are related to normalization constants of
bound-state wavefunctions (see, e.g., \cite[\S128]{Landau:1991wop}),
also satisfy the same geometric scaling.

The above discussion, although simplified, captures the general impact
of discrete scale invariance in quantum theory. To date, there have
been discovered a number of quantum systems that enjoy discrete scale
invariance, log-periodicity, or geometric scaling; see
\cite{Ovdat:2019ywg} for a nice review. Among the notable examples is
the Efimov effect \cite{Efimov:1970zz,Efimov:1973awb} in three-body
problems under two-body short-range interactions, where there emerge
the geometric series of three-body bound states if scattering lengths
diverge and dimensionful parameters apparently disappear. Note,
however, that this Efimov effect is known to be highly susceptible to
particle statistics and dimensionality. For example, for three
identical bosons, it was shown that the Efimov effect is present only
when the spatial dimension $d$ is in the range $2.3<d<3.8$
\cite{Nielsen:2001}. As discussed in \cite{Nishida:2011ew}, this was
due to the absence of nontrivial scale-invariant two-body contact
interactions (at least in the limit of infinite scattering length) in
other dimensions.

One purpose of this paper is to show that---contrary to the
conventional wisdom---there in fact exist a lot of scale-invariant
two-body contact interactions in one dimension if the number of
particles is greater than two. Another purpose is then to present
concrete examples of one-dimensional $n(\geq3)$-body problems that
undergo phase transition from a continuous scale-invariant phase to a
discrete scale-invariant phase. For the sake of simplicity, in this
work we focus on identical spinless particles that interact only
through two-body contacts.\footnote{For other systems that realize
  discrete scale invariance in one dimension, see
  \cite{Nishida:2009pg,Nishida:2011ew,Moroz:2015}.} Remarkably, any
such many-body systems generally enjoy the \textit{boson-fermion
  duality}---the one-to-one correspondence between isospectral bosonic
and fermionic systems---which enables us to treat bosons and fermions
on equal footing. In essence, the boson-fermion duality in one
dimension is just the equivalence between the even-parity sector of
the $\delta$-function potential system and the odd-parity sector of
the $\varepsilon$-function potential system \cite{Cheon:2000tq}. The
simplest application of this equivalence to many-body problems is the
well-known boson-fermion duality between the Lieb-Liniger model
\cite{Lieb:1963rt} of identical spinless bosons and the
Cheon-Shigehara model \cite{Cheon:1998iy} of identical spinless
fermions. Recently, it has been shown \cite{Ohya:2021oni} that this
duality can be further generalized because one-dimensional two-body
contact interactions have much more variety than previously
investigated. And most importantly, this generalization includes
scale-invariant two-body contact interactions which---at least at the
formal level---render the system invariant under continuous scale
transformation. Such continuous scale invariance, however, can be
broken down to discrete scale invariance just as in the Efimov
effect. The goal of this paper is to show that this indeed happens for
both bosons and fermions and to present the \textit{exact} $n$-body
bound-state spectrum as well as the \textit{exact} $n$-body S-matrix
elements that exhibit geometric scaling and log-periodicity. The key
to this achievement is the \textit{configuration-space approach} to
identical particles
\cite{Souriau:1967,Souriau:1969,Laidlaw:1970ei,Leinaas:1977fm}. Before
going to discuss scale-invariance breaking, let us first briefly
review the boson-fermion duality in \cite{Ohya:2021oni} from the
viewpoint of the configuration-space approach.

\section{Boson-fermion duality in one dimension}
\label{section:2}
\begin{figure}[t]
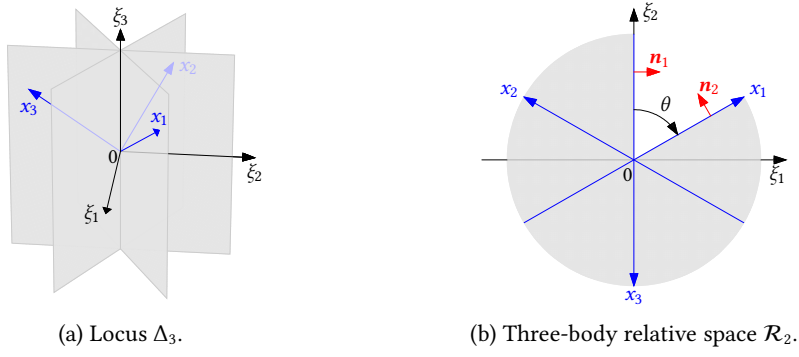

  \centering%
  \begin{subfigure}[t]{0.4\linewidth}
    \centering%
    \input{figure1a.eepic}%
    \caption{Locus $\Delta_{3}$.}
    \label{figure:1a}
  \end{subfigure}\quad
  \centering%
  \begin{subfigure}[t]{0.4\linewidth}
    \centering%
    \input{figure1b.eepic}%
    \caption{Three-body relative space $\mathcal{R}_{2}$.}
    \label{figure:1b}
  \end{subfigure}
  \caption{Configuration space of three identical particles in one
    dimension. (a) The gray-shaded regions represent the locus
    $\Delta_{3}=\{(x_{1},x_{2},x_{3}):(x_{1}-x_{2})(x_{1}-x_{3})(x_{2}-x_{3})=0\}$
    which splits $\mathbb{R}^{3}$ into $3!$ disconnected regions. The
    three-body configuration space
    $\mathcal{M}_{3}=\{(x_{1},x_{2},x_{3}):x_{1}>x_{2}>x_{3}\}$ is
    just one of those disconnected regions. The $\xi_{1}$-,
    $\xi_{2}$-, and $\xi_{3}$-axes are pointing along the directions
    of the unit vectors $\bm{e}_{1}=\frac{1}{\sqrt{2}}(1,-1,0)$,
    $\bm{e}_{2}=\frac{1}{\sqrt{6}}(1,1,-2)$, and
    $\bm{e}_{3}=\frac{1}{\sqrt{3}}(1,1,1)$. (b) The blank white region
    represents the relative space
    $\mathcal{R}_{2}=\{(\xi_{1},\xi_{2}):0<\xi_{1}<\sqrt{3}\xi_{2}\}$
    which is just the $\xi_{3}=\text{const}$ section of
    $\mathcal{M}_{3}$. The gray-shaded region represents the
    impenetrable region for identical particles. The red arrows
    represent the inward-pointing unit normal vectors
    $\bm{n}_{1}=\frac{1}{\sqrt{2}}(1,-1,0)$ and
    $\bm{n}_{2}=\frac{1}{\sqrt{2}}(0,1,-1)$.}
  \label{figure:1}
\end{figure}

Roughly speaking, the configuration-space approach is an approach to
identical particles where permutation invariance is regarded as
\textit{gauge symmetry}; that is, invariance of physical observables
under permutation of multiparticle coordinates is merely a redundancy
in description \cite{Leinaas:1977fm}. As in any gauge theory, every
gauge-equivalent configurations are physically equivalent such that
the configuration space must be a collection of inequivalent gauge
orbits. To be more precise, given a one-particle configuration space
$X$, the $n$-particle configuration space of identical particles is
generally given by the orbit space
$\mathcal{M}_{n}=\mathring{X}^{n}/S_{n}$, where
$\mathring{X}^{n}=X^{n}-\Delta_{n}$ is the configuration space of $n$
distinguishable particles and $S_{n}$ the symmetric group. Here
$X^{n}$ stands for the Cartesian product of $n$ copies of $X$ and
$\Delta_{n}$ the set of coincidence points at which two or more
particles occupy the same place simultaneously. In general, such a set
can be defined as the following locus:
\begin{align}
  \Delta_{n}=\left\{(x_{1},\cdots,x_{n})\in X^{n}:\prod_{1\leq j<k\leq n}\|x_{j}-x_{k}\|=0\right\},\label{eq:3}
\end{align}
where $\|\cdot\|$ stands for the norm equipped with $X^{n}$. Note that
many-body contact interactions are those that have support only on
$\Delta_{n}$, where wavefunctions become singular in general. Figure
\ref{figure:1a} shows $\Delta_{3}$ for $X=\mathbb{R}$. (Note that, for
$X=\mathbb{R}$, $\Delta_{n}$ can also be defined as the vanishing
locus of the Vandermonde polynomial,
$\prod_{1\leq j<k\leq n}(x_{j}-x_{k})=0$.)

Now let us focus on the case $X=\mathbb{R}$, in which
$\mathring{\mathbb{R}}^{n}\ni(x_{1},\cdots,x_{n})$ consists of $n!$
disconnected regions described by the inequality
$x_{\sigma(1)}>\cdots>x_{\sigma(n)}$, where $\sigma\in S_{n}$ is a
permutation of $n$ indices. All of these $n!$ regions are gauge
equivalent for identical particles. Hence the configuration space of
$n$ identical particles in one dimension can be identified with the
following $n$-dimensional space:
\begin{align}
  \mathcal{M}_{n}=\{(x_{1},\cdots,x_{n}):x_{1}>\cdots>x_{n}\}.\label{eq:4}
\end{align}
Note that this space has a number of nontrivial boundaries; see
Fig.~\ref{figure:1b} for the case $n=3$. Of particular importance are
the following codimension-1 boundaries at which two out of $n$
particles collide:
\begin{align}
  \partial\mathcal{M}_{n,j}^{\text{2-body}}=\{(x_{1},\cdots,x_{n}):x_{1}>\cdots>x_{j}=x_{j+1}>\cdots>x_{n}\},\label{eq:5}
\end{align}
where $j=1,\cdots,n-1$.

Let us now focus on the situation where identical particles freely
propagate almost everywhere on the line yet interact only at the
two-body coincidence points. Since all the coincidence points are
excluded in $\mathcal{M}_{n}$, in the configuration-space approach the
$n$-body Hamiltonian for such systems is just the following free
Hamiltonian:
\begin{align}
  H_{0}=-\frac{\hbar^{2}}{2m}\sum_{j=1}^{n}\frac{\partial^{2}}{\partial x_{j}^{2}},\label{eq:6}
\end{align}
where $m$ is the mass of the identical particles. The two-body contact
interactions are then described by boundary conditions of
wavefunctions at the codimension-1 boundaries \eqref{eq:5}. Such
boundary conditions must be chosen to fulfill unitarity, or
probability conservation. It is well known that such boundary
conditions are generally given by the following Robin boundary
conditions:
\begin{align}
  \frac{\partial\psi}{\partial n_{j}}-\frac{1}{a_{j}}\psi=0\quad\text{on}\quad\partial\mathcal{M}_{n,j}^{\text{2-body}},\label{eq:7}
\end{align}
where $\partial\psi/\partial n_{j}$ stands for the normal derivative
to the boundary $\partial\mathcal{M}_{n,j}^{\text{2-body}}$ given by
\begin{align}
  \frac{\partial\psi}{\partial n_{j}}
  =\bm{n}_{j}\cdot\bm{\nabla}\psi
  =\frac{1}{\sqrt{2}}\left(\frac{\partial}{\partial x_{j}}-\frac{\partial}{\partial x_{j+1}}\right)\psi.\label{eq:8}
\end{align}
Here
$\bm{n}_{j}=\bm{\nabla}(x_{j}-x_{j+1})/\|\bm{\nabla}(x_{j}-x_{j+1})\|=\frac{1}{\sqrt{2}}(0,\cdots,0,1,-1,0,\cdots,0)$
is the inward-pointing unit normal vector\footnote{Note that the
  normalization is different from the previous work
  \cite{Ohya:2021oni} where we have chosen $\|\bm{n}_{j}\|=\sqrt{2}$.}
to the surface $x_{j}-x_{j+1}=0$,
$\bm{\nabla}=(\frac{\partial}{\partial
  x_{1}},\cdots,\frac{\partial}{\partial x_{n}})$ is the derivative on
$\mathcal{M}_{n}$, and $a_{j}$ is a real parameter that can depend on
the coordinates orthogonal to $\bm{n}_{j}$. In this way, in the
configuration-space approach the free Hamiltonian \eqref{eq:6} and the
Robin boundary conditions \eqref{eq:7} set the problem of identical
spinless particles under two-body contact interactions.

Now, one may want to know how Eqs.~\eqref{eq:6} and \eqref{eq:7}
describe the boson-fermion duality in the conventional approach, where
the configuration space is taken to be $\mathring{\mathbb{R}}^{n}$
rather than $\mathring{\mathbb{R}}^{n}/S_{n}$. To see this, let us
first construct conventional bosonic and fermionic wavefunctions on
$\mathring{\mathbb{R}}^{n}$, which can easily be done by extending the
domain of wavefunctions. Let $\psi$ be a normalized wavefunction on
$\mathcal{M}_{n}$ and let $\bm{x}=(x_{1},\cdots,x_{n})$ be in the
region $x_{\sigma(1)}>\cdots>x_{\sigma(n)}$. Then we define
\begin{subequations}
  \begin{align}
    \psi_{\text{B}}(\bm{x})&=\frac{1}{\sqrt{n!}}\psi(\sigma\bm{x}),\label{eq:9a}\\
    \psi_{\text{F}}(\bm{x})&=\frac{1}{\sqrt{n!}}\operatorname{sgn}(\sigma)\psi(\sigma\bm{x}),\label{eq:9b}
  \end{align}
\end{subequations}
where $\sigma\bm{x}=(x_{\sigma(1)},\cdots,x_{\sigma(n)})$ and
$\operatorname{sgn}(\sigma)$ stands for the signature of
$\sigma\in S_{n}$. As $\sigma$ runs through all the permutations,
Eqs.~\eqref{eq:9a} and \eqref{eq:9b} define the totally symmetric and
antisymmetric functions on $\mathring{\mathbb{R}}^{n}$, thus providing
wavefunctions of identical spinless bosons and fermions in the
conventional approach. By construction, it is obvious that there holds
the identity
$\psi_{\text{F}}(\bm{x})=\sgn(\sigma)\psi_{\text{B}}(\bm{x})$ in the
region $x_{\sigma(1)}>\cdots>x_{\sigma(n)}$, which can be extended to
$\mathring{\mathbb{R}}^{n}$ in the following way:
\begin{align}
  \psi_{\text{F}}(\bm{x})=\left(\prod_{1\leq j<k\leq n}\sgn(x_{j}-x_{k})\right)\psi_{\text{B}}(\bm{x}),\quad\forall\bm{x}\in\mathring{\mathbb{R}}^{n},\label{eq:10}
\end{align}
where $\sgn(x)=x/|x|$ stands for the sign function. In this way, for
identical spinless particles, there holds the one-to-one
correspondence between the bosonic and fermionic wavefunctions in the
conventional approach. This is the celebrated boson-fermion mapping in
one dimension \cite{Girardeau:1960}.

Let us next construct the Hamiltonians for $\psi_{\text{B}}$ and
$\psi_{\text{F}}$, which can be achieved by studying connection
conditions at the codimension-1 singularities in
$\mathring{\mathbb{R}}^{n}$. To this end, let us first start with the
following toy example:
\begin{align}
  f^{\prime}(0_{+})-\frac{1}{a}f(0_{+})=0,\label{eq:11}
\end{align}
where $f(x)$ is some function on $\mathbb{R}$ and the prime indicates
the derivative with respect to $x$. If $f(x)$ is an even function that
satisfies $f(x)=f(-x)$ and $f^{\prime}(x)=-f^{\prime}(-x)$, there
automatically hold $f(0_{+})=f(0_{-})$ and
$f^{\prime}(0_{+})=-f^{\prime}(0_{-})$. Hence, for such even
functions, the boundary condition \eqref{eq:11} is equivalent to the
connection condition
$f^{\prime}(0_{+})-f^{\prime}(0_{-})-\frac{1}{a}(f(0_{+})+f(0_{-}))=0$
at $x=0$. On the other hand, if $f(x)$ is an odd function that
satisfies $f(x)=-f(-x)$ and $f^{\prime}(x)=f^{\prime}(-x)$, there
automatically hold $f(0_{+})=-f(0_{-})$ and
$f^{\prime}(0_{+})=f^{\prime}(0_{-})$. Hence, for such odd functions,
the boundary condition \eqref{eq:11} is equivalent to the connection
condition
$f^{\prime}(0_{+})+f^{\prime}(0_{-})-\frac{1}{a}(f(0_{+})-f(0_{-}))=0$.

The above discussion can easily be generalized to $\psi_{\text{B}}$
and $\psi_{\text{F}}$. A careful analysis shows that, for totally
symmetric functions, the Robin boundary condition \eqref{eq:7} is
equivalent to the following connection condition at the codimension-1
singularity
$\{x_{\sigma(1)}>\cdots>x_{\sigma(j)}=x_{\sigma(j+1)}>\cdots>x_{\sigma(n)}\}$
in $\mathring{\mathbb{R}}^{n}$ \cite{Ohya:2021oni}:
\begin{align}
  \left(\frac{\partial}{\partial x_{\sigma(j)}}-\frac{\partial}{\partial x_{\sigma(j+1)}}\right)\psi_{\text{B}}\biggr|_{0_{+}}
  -\left(\frac{\partial}{\partial x_{\sigma(j)}}-\frac{\partial}{\partial x_{\sigma(j+1)}}\right)\psi_{\text{B}}\biggr|_{0_{-}}
  -\frac{\sqrt{2}}{a_{j}}\left(\psi_{\text{B}}\bigr|_{0_{+}}+\psi_{\text{B}}\bigr|_{0_{-}}\right)=0,\label{eq:12}
\end{align}
where $|_{0_{\pm}}$ is shorthand for
$|_{x_{\sigma(j)}-x_{\sigma(j+1)}=0_{\pm}}$ and $\sigma$ is an even
permutation. For totally antisymmetric functions, on the other hand,
the Robin boundary condition \eqref{eq:7} is equivalent to the
following connection condition:
\begin{align}
  \left(\frac{\partial}{\partial x_{\sigma(j)}}-\frac{\partial}{\partial x_{\sigma(j+1)}}\right)\psi_{\text{F}}\biggr|_{0_{+}}
  +\left(\frac{\partial}{\partial x_{\sigma(j)}}-\frac{\partial}{\partial x_{\sigma(j+1)}}\right)\psi_{\text{F}}\biggr|_{0_{-}}
  -\frac{\sqrt{2}}{a_{j}}\left(\psi_{\text{F}}\bigr|_{0_{+}}-\psi_{\text{F}}\bigr|_{0_{-}}\right)=0.\label{eq:13}
\end{align}
Note that, thanks to the symmetry property, $\psi_{\text{B}}$ and the
normal derivative of $\psi_{\text{F}}$ are both continuous at the
codimension-1 singularities.

Now, Eq.~\eqref{eq:12} together with
$\psi_{\text{B}}|_{0_{+}}=\psi_{\text{B}}|_{0_{-}}$ is nothing but the
connection condition for the $\delta$-function potential
$\delta(x_{\sigma(j)}-x_{\sigma(j+1)};\tfrac{\sqrt{2}}{a_{j}})=\tfrac{\sqrt{2}}{a_{j}}\delta(x_{\sigma(j)}-x_{\sigma(j+1)})$
supported on the codimension-1 singularity
$\{x_{\sigma(1)}>\cdots>x_{\sigma(j)}=x_{\sigma(j+1)}>\cdots>x_{\sigma(n)}\}$
in $\mathring{\mathbb{R}}^{n}$, where the coupling constant is
$\sqrt{2}/a_{j}$. On the other hand, Eq.~\eqref{eq:13} together with
$(\frac{\partial}{\partial x_{\sigma(j)}}-\frac{\partial}{\partial
  x_{\sigma(j+1)}})\psi_{\text{F}}|_{0_{+}}=(\frac{\partial}{\partial
  x_{\sigma(j)}}-\frac{\partial}{\partial
  x_{\sigma(j+1)}})\psi_{\text{F}}|_{0_{-}}$ is nothing but the
connection conditions for the $\varepsilon$-function
potential\footnote{The $\varepsilon$-function potential is defined by
  $\varepsilon(x;c)=\lim_{a\to0}(\frac{1}{2c}-\frac{1}{2a})(\delta(x+a)+\delta(x-a))$
  and is described by the connection conditions
  $\psi(0_{+})-\psi(0_{-})-c(\psi^{\prime}(0_{+})+\psi^{\prime}(0_{-}))=0$
  and $\psi^{\prime}(0_{+})=\psi^{\prime}(0_{-})$
  \cite{Cheon:1998iy}. (See also \cite{Cheon:1997rx} for another
  definition of $\varepsilon(x;c)$.) Note that the contact interaction
  described by these connection conditions is also known as the
  ``$\delta^{\prime}$-interaction'' in the literature; see, e.g.,
  Chapter I.4 of \cite{Albeverio:1988}.}
$\varepsilon(x_{\sigma(j)}-x_{\sigma(j+1)};\tfrac{a_{j}}{\sqrt{2}})$,
where in this case the coupling constant is $a_{j}/\sqrt{2}$. Thus we
find the following Hamiltonians for $\psi_{\text{B}}$ and
$\psi_{\text{F}}$:
\begin{align}
  H_{\text{B}/\text{F}}=H_{0}+V_{\text{B}/\text{F}},\label{eq:14}
\end{align}
where
\begin{subequations}
  \begin{align}
    V_{\text{B}}&=\frac{\hbar^{2}}{m}\sum_{j=1}^{n-1}\sum_{\sigma\in A_{n}}\left[\prod_{k\in\{1,\cdots,n-1\}\setminus\{j\}}\theta(x_{\sigma(k)}-x_{\sigma(k+1)})\right]\delta(x_{\sigma(j)}-x_{\sigma(j+1)};\tfrac{\sqrt{2}}{a_{j}}),\label{eq:15a}\\
    V_{\text{F}}&=\frac{\hbar^{2}}{m}\sum_{j=1}^{n-1}\sum_{\sigma\in A_{n}}\left[\prod_{k\in\{1,\cdots,n-1\}\setminus\{j\}}\theta(x_{\sigma(k)}-x_{\sigma(k+1)})\right]\varepsilon(x_{\sigma(j)}-x_{\sigma(j+1)};\tfrac{a_{j}}{\sqrt{2}}).\label{eq:15b}
  \end{align}
\end{subequations}
Here $A_{n}$ is the alternating group that consists of only even
permutations. The factor $\prod\theta(x_{\sigma(k)}-x_{\sigma(k+1)})$
is introduced in order to guarantee the ordering
$x_{\sigma(1)}>\cdots>x_{\sigma(j)}=x_{\sigma(j+1)}>\cdots>x_{\sigma(n)}$,
where $\theta(x)$ is the step function. Note that, since the coupling
constants of the two systems are inverse to each other, there holds
the one-to-one correspondence between the strong-coupling regime in
one system and the weak-coupling regime in the other. This is a
natural generalization of the celebrated strong-weak duality in
\cite{Cheon:1998iy}. Note also that, since the eigenvalue equations
$H_{\text{B}}\psi_{\text{B}}=E\psi_{\text{B}}$ and
$H_{\text{F}}\psi_{\text{F}}=E\psi_{\text{F}}$ both boil down to
$H_{0}\psi=E\psi$ on $\mathcal{M}_{n}$ with the same boundary
conditions, $H_{\text{B}}$ and $H_{\text{F}}$ are completely
isospectral.

To summarize, the $n$-body problem described by the free Hamiltonian
\eqref{eq:6} and the Robin boundary conditions \eqref{eq:7} on
$\mathcal{M}_{n}$ is equivalent to the $n$-boson and $n$-fermion
problems described by $H_{\text{B}}$ and $H_{\text{F}}$. By
construction, we have the spectral equivalence between $H_{\text{B}}$
and $H_{\text{F}}$, the boson-fermion mapping between
$\psi_{\text{B}}$ and $\psi_{\text{F}}$, and the strong-weak
duality. Notice that, if $a_{1}=\cdots=a_{n-1}=\text{const}$,
Eq.~\eqref{eq:14} just reduces to the Lieb-Liniger model and the
Cheon-Shigehara model. Note also that, since the $n$-body Hamiltonian
\eqref{eq:14} is of the form
$H=-\frac{\hbar^{2}}{2m}\sum_{j=1}^{n}\frac{\partial^{2}}{\partial
  x_{j}^{2}}+V(x_{1},\cdots,x_{n})$, it in general does not admit any
nontrivial cluster decomposition into the sum of cluster Hamiltonians
and intercluster potentials. In other words, the $n$-boson and
$n$-fermion systems in the present paper are generally $n$-body
clusters that cannot be decomposed into subclusters. We will elaborate
on this cluster property in Sec.~\ref{section:5} by using the
three-body scattering problem.

Now, as noted before, $a_{j}$ can depend on the coordinates orthogonal
to $\bm{n}_{j}$ without spoiling unitarity. This opens up a new vista
for realizing scale invariance in one-dimensional $n$-body problems
under two-body contact interactions. Let us next move on to study such
a scale-invariant subfamily of the Robin boundary conditions
\eqref{eq:7}.

\section{Scale-invariant two-body boundary conditions}
\label{section:3}
To begin with, let us first introduce the normalized Jacobi
coordinates $(\xi_{1},\cdots,\xi_{n})$ in $\mathcal{M}_{n}$, which can
be defined through the orthogonal transformation
$x_{j}\mapsto\xi_{j}=\bm{e}_{j}\cdot\bm{x}=\sum_{k=1}^{n}e_{jk}x_{k}$. Here
$\bm{e}_{j}=(e_{j1},\cdots,e_{jn})$ is the following $n$-dimensional
orthonormal vector:
\begin{subequations}
  \begin{align}
    \bm{e}_{j}&=\frac{1}{\sqrt{j(j+1)}}(1,\cdots,1,-j,0,\cdots,0),\quad j\in\{1,\cdots,n-1\},\label{eq:16a}\\
    \bm{e}_{n}&=\frac{1}{\sqrt{n}}(1,\cdots,1),\label{eq:16b}
  \end{align}
\end{subequations}
where $-j$ in Eq.~\eqref{eq:16a} is in the $(j+1)$th component. Note
that $\bm{x}$ can be written as
$\bm{x}=\xi_{1}\bm{e}_{1}+\cdots+\xi_{n}\bm{e}_{n}$. Note also that
$\xi_{n}=\frac{x_{1}+\cdots+x_{n}}{\sqrt{n}}$, which ranges from
$-\infty$ to $\infty$, corresponds to the center-of-mass
coordinates. Hence it is convenient to separate the one-dimensional
subspace
$\mathbb{R}\bm{e}_{n}=\{\xi_{n}\bm{e}_{n}:-\infty<\xi_{n}<\infty\}$
from $\mathcal{M}_{n}$. It is also convenient to introduce the
hyperspherical coordinates in the $(n-1)$-dimensional subspace spanned
by the set of vectors $\{\bm{e}_{1},\cdots,\bm{e}_{n-1}\}$, which
describes the relative space $\mathcal{R}_{n-1}$ (the configuration
space of relative motion). First, the hyperradius in the relative
space $\mathcal{R}_{n-1}$ is defined by
\begin{align}
  r
  &=\sqrt{\xi_{1}^{2}+\cdots+\xi_{n-1}^{2}}\nonumber\\
  &=\sqrt{\frac{1}{n}\sum_{1\leq j<k\leq n}(x_{j}-x_{k})^{2}}.\label{eq:17}
\end{align}
The hyperangles in the relative space $\mathcal{R}_{n-1}$ are then
described by the unit vector
$\Hat{\bm{\xi}}=(\Hat{\xi}_{1},\cdots,\Hat{\xi}_{n-1}):=\frac{1}{r}(\xi_{1},\cdots,\xi_{n-1})$,
which, from the condition $x_{1}>\cdots>x_{n}$, must satisfy the
condition
$0<\Hat{\xi}_{1}<\cdots<\sqrt{\frac{n(n-1)}{2}}\Hat{\xi}_{n-1}$
\cite{Ohya:2021oni}. The configuration space is then factorized as
follows:
\begin{align}
  \mathcal{M}_{n}=\mathbb{R}\times\mathbb{R}_{+}\times\Omega_{n-2},\label{eq:18}
\end{align}
where $\mathbb{R}=\{\xi_{n}:-\infty<\xi_{n}<\infty\}$ is the space of
the center-of-mass motion, $\mathbb{R}_{+}=\{r:0<r<\infty\}$ is the
space of the hyperradial motion, and $\Omega_{n-2}$ is the space of
the hyperangular motion given by
\begin{align}
  \Omega_{n-2}=\Bigl\{(\Hat{\xi}_{1},\cdots,\Hat{\xi}_{n-1}):\Hat{\xi}_{1}^{2}+\cdots+\Hat{\xi}_{n-1}^{2}=1,~0<\Hat{\xi}_{1}<\cdots<\sqrt{\tfrac{n(n-1)}{2}}\Hat{\xi}_{n-1}\Bigr\}.\label{eq:19}
\end{align}
As we will see shortly, this factorization plays a pivotal role in
solving the $n$-body Schr\"{o}dinger equation by the method of
separation of variables. Note that, for $n=2$, the factor
$\Omega_{n-2}$ should be discarded in Eq.~\eqref{eq:18}. Note also
that the relative space
$\mathcal{R}_{n-1}=\mathbb{R}_{+}\times\Omega_{n-2}$ can also be
written as
$\mathcal{R}_{n-1}=\{(\xi_{1},\cdots,\xi_{n}):0<\xi_{1}<\cdots<\sqrt{\frac{n(n-1)}{2}}\xi_{n-1}\}$.

Now, in the coordinate system $(\xi_{n},r,\Hat{\bm{\xi}})$, the
gradient of a wavefunction $\psi$ is written as follows:
\begin{align}
  \bm{\nabla}\psi=\frac{\partial\psi}{\partial\xi_{n}}\bm{e}_{\xi_{n}}+\frac{\partial\psi}{\partial r}\bm{e}_{r}+\frac{1}{r}\bm{\nabla}_{\Omega_{n-2}}\psi,\label{eq:20}
\end{align}
where $\bm{e}_{\xi_{n}}$ and $\bm{e}_{r}$ are the unit vectors
pointing along the $\xi_{n}$- and $r$-directions and
$\bm{\nabla}_{\Omega_{n-2}}$ is the gradient on $\Omega_{n-2}$. Notice
that, for $n\geq3$, all the normal vectors $\bm{n}_{j}$ are orthogonal
to $\bm{e}_{\xi_{n}}$ and $\bm{e}_{r}$. (For $n=2$, the normal vector
is equivalent to $\bm{e}_{r}$.) Hence, for $n\geq3$, the Robin
boundary condition \eqref{eq:7} is cast into the following
form:\footnote{For $n=2$, Eq.~\eqref{eq:7} becomes
  $\frac{\partial\psi}{\partial r}-\frac{1}{a_{1}}\psi=0$ at $r=0$. In
  this case there is no scale- and translation-invariant boundary
  condition except for the trivial Dirichlet and Neumann boundary
  conditions.}
\begin{align}
  \frac{1}{r}\bm{n}_{j}\cdot\bm{\nabla}_{\Omega_{n-2}}\psi-\frac{1}{a_{j}}\psi=0\quad\text{on}\quad\partial\mathcal{M}_{n,j}^{\text{2-body}}.\label{eq:21}
\end{align}
Below we will focus on the case $n\geq3$.

Now we are ready to identify the scale-invariant subfamily of the
boundary conditions. First of all, under the scale transformation
$\mathcal{S}_{\alpha}:\bm{x}\mapsto\alpha\bm{x}$ ($\alpha>0$),
one-dimensional $n$-body wavefunctions transform as follows:
\begin{align}
  \psi(\bm{x})\mapsto(\mathcal{S}_{\alpha}\psi)(\bm{x}):=\alpha^{\frac{n}{2}}\psi(\alpha\bm{x}).\label{eq:22}
\end{align}
The boundary condition \eqref{eq:21} is then said to be scale
invariant if $\mathcal{S}_{\alpha}\psi$ satisfies the same boundary
condition as $\psi$. It is, however, obvious from the factor
$\frac{1}{r}$ that Eq.~\eqref{eq:21} does not remain unchanged under
$\mathcal{S}_{\alpha}$ unless $a_{j}$ depends on the coordinates and
transforms as
$a_{j}(\bm{x})\mapsto a_{j}(\alpha\bm{x})=\alpha a_{j}(\bm{x})$. (Note
that $\bm{\nabla}_{\Omega_{n-2}}$ is invariant under
$\mathcal{S}_{\alpha}$.) Note also that coordinate-dependent $a_{j}$
generally breaks translation invariance unless it satisfies
$a_{j}(x_{1}+\beta,\cdots,x_{n}+\beta)=a_{j}(x_{1},\cdots,x_{n})$ for
any real $\beta$. Hence, in order to realize scale- and
translation-invariant boundary conditions, $a_{j}(\bm{x})$ must
satisfy the following conditions:
\begin{subequations}
  \begin{align}
    \text{scaling law:}\quad&a_{j}(\alpha\bm{x})=\alpha a_{j}(\bm{x}),\label{eq:23a}\\
    \text{translation invariance:}\quad&a_{j}(\bm{x}+\beta\bm{e}_{n})=a_{j}(\bm{x}).\label{eq:23b}
  \end{align}
\end{subequations}
It is easy to see that the general solution to these conditions is
given by
\begin{align}
  a_{j}(\bm{x})=rg_{j}(\bm{\Hat{\xi}}),\label{eq:24}
\end{align}
where $g_{j}$ is an arbitrary function of $\bm{\Hat{\xi}}$. The
boundary condition \eqref{eq:21} then becomes
\begin{align}
  \bm{n}_{j}\cdot\bm{\nabla}_{\Omega_{n-2}}\psi-\frac{1}{g_{j}}\psi=0\quad\text{on}\quad\partial\mathcal{M}_{n,j}^{\text{2-body}}.\label{eq:25}
\end{align}
This describes the most general scale- and translation-invariant
two-body contact interactions in the $n(\geq3)$-body problems of
identical spinless particles in one dimension. Since both the
Hamiltonian and the boundary conditions are scale invariant, the
$n$-body system described by Eqs.~\eqref{eq:6} and \eqref{eq:25}
is---at least formally---scale invariant. This continuous scale
invariance, however, can be broken down to discrete scale invariance
in exactly the same way as the Efimov effect. Let us next investigate
the criterion for such symmetry breaking by using the $n$-body
Schr\"{o}dinger equation.

\section{From continuous to discrete scale invariance}
\label{section:4}
Let us study the time-independent $n$-body Schr\"{o}dinger equation
$H_{0}\psi=E\psi$. To this end, we first note that, in the coordinate
system $(\xi_{n},r,\Hat{\bm{\xi}})$, the free Hamiltonian \eqref{eq:6}
can be written as
$H_{0}=r^{-\frac{n-2}{2}}\widetilde{H}_{0}r^{\frac{n-2}{2}}$,
where\footnote{Essentially the same Hamiltonian as in
  Eq.~\eqref{eq:26} was discussed in
  \cite{Dehkharghani:2015,Harshman:2017} in the context of
  nonidentical particles.}
\begin{align}
  \widetilde{H}_{0}=-\frac{\hbar^{2}}{2m}\left(\frac{\partial^{2}}{\partial r^{2}}+\frac{\Delta_{\Omega_{n-2}}-\frac{(n-2)(n-4)}{4}}{r^{2}}+\frac{\partial^{2}}{\partial\xi_{n}^{2}}\right).\label{eq:26}
\end{align}
Here $\Delta_{\Omega_{n-2}}$ stands for the Laplacian on
$\Omega_{n-2}$. Hence, under the assumption that the wavefunction has
the form
\begin{align}
  \psi(\bm{x})=r^{-\frac{n-2}{2}}\Psi(\xi_{n})R(r)\Theta(\Hat{\bm{\xi}}),\label{eq:27}
\end{align}
the Schr\"{o}dinger equation $H_{0}\psi=E\psi$ boils down to the
following differential equations:
\begin{subequations}
  \begin{align}
    -\frac{\partial^{2}}{\partial\xi_{n}^{2}}\Psi(\xi_{n})&=\frac{2mE_{\text{cm}}}{\hbar^{2}}\Psi(\xi_{n}),\label{eq:28a}\\
    -\Delta_{\Omega_{n-2}}\Theta(\Hat{\bm{\xi}})&=\lambda\Theta(\Hat{\bm{\xi}}),\label{eq:28b}\\
    \left(-\frac{\partial^{2}}{\partial r^{2}}+\frac{\lambda+\frac{(n-2)(n-4)}{4}}{r^{2}}\right)R(r)&=\frac{2m E_{\text{rel}}}{\hbar^{2}}R(r),\label{eq:28c}
  \end{align}
\end{subequations}
where $E_{\text{cm}}+E_{\text{rel}}=E$. Note that the boundary
condition \eqref{eq:25} is only for the hyperangular wavefunction
$\Theta$. In other words, all the information about the two-body
contact interactions is encoded in the eigenvalue $\lambda$. Note also
that, since $\xi_{n}$ and $r$ are permutation invariant, the totally
symmetric and antisymmetric wavefunctions in
$\mathring{\mathbb{R}}^{n}$ are obtained by just extending the domain
of $\Theta$. For example, in the region where $\xi_{n}\in\mathbb{R}$,
$r\in\mathbb{R}_{+}$, and $\sigma\Hat{\bm{\xi}}\in\Omega_{n-2}$, where
$\sigma\Hat{\bm{\xi}}$ stands for the action of the permutation
$\sigma\in S_{n}$ on the unit vector $\Hat{\bm{\xi}}$, we have
\begin{subequations}
  \begin{align}
    \psi_{\text{B}}(\xi_{n},r,\Hat{\bm{\xi}})&=\frac{1}{\sqrt{n!}}r^{-\frac{n-2}{2}}\Psi(\xi_{n})R(r)\Theta(\sigma\Hat{\bm{\xi}}),\label{eq:29a}\\
    \psi_{\text{F}}(\xi_{n},r,\Hat{\bm{\xi}})&=\frac{1}{\sqrt{n!}}\sgn(\sigma)r^{-\frac{n-2}{2}}\Psi(\xi_{n})R(r)\Theta(\sigma\Hat{\bm{\xi}}).\label{eq:29b}
  \end{align}
\end{subequations}
As $\sigma$ runs through all the permutations, the above equations
define the $n$-body wavefunctions of identical bosons and fermions in
$\mathring{\mathbb{R}}^{n}$. It should be emphasized that, if $\Psi$,
$R$, and $\Theta$ are normalized solutions to
Eqs.~\eqref{eq:28a}--\eqref{eq:28c}, then Eqs.~\eqref{eq:29a} and
\eqref{eq:29b} automatically become the normalized eigenfunctions of
the Hamiltonians \eqref{eq:14} with the eigenvalue
$E=E_{\text{cm}}+E_{\text{rel}}$.

Now, it is well known in the context of the $1/r^{2}$ potential that
infinitely many discrete energy levels appear if
$\lambda+\frac{(n-2)(n-4)}{4}<-\frac{1}{4}$ \cite{Case:1950an}; that
is, continuous scale invariance is broken down to discrete scale
invariance if $\lambda<\lambda_{\text{c}}$, where
$\lambda_{\text{c}}=-\frac{(n-3)^{2}}{4}$ is the critical
value.\footnote{For
  $-\frac{1}{4}<\lambda+\frac{(n-2)(n-4)}{4}<\frac{3}{4}$, there
  exists a one-parameter family of self-adjoint extensions of the
  Hamiltonian. For simplicity, we will not discuss this issue in the
  present paper.} Hence the sufficient condition for the
scale-invariance breaking is
\begin{align}
  \inf\sigma(-\Delta_{\Omega_{n-2}})<\lambda_{\text{c}},\label{eq:30}
\end{align}
where $\sigma(-\Delta_{\Omega_{n-2}})$ here stands for the spectrum of
the operator $-\Delta_{\Omega_{n-2}}$. Let us, for the moment, assume
that there exists at least one such negative eigenvalue and see the
impact of discrete scale invariance by solving the hyperradial
equation \eqref{eq:28c} exactly.

\subsection{Exact \texorpdfstring{\boldmath$n$}{n}-body bound-state
  spectrum}
\label{section:4A}
Let us first consider the case $E_{\text{rel}}<0$. In this case the
normalized solution to the hyperradial equation \eqref{eq:28c} for
$\lambda<\lambda_{\text{c}}$ is given by
\begin{align}
  R_{\kappa\lambda}(r)
  &=N_{\kappa}\sqrt{\frac{2\kappa r}{\pi}}K_{i\nu}(\kappa r)\nonumber\\
  &\to N_{\kappa}\e^{-\kappa r}\quad\text{as}\quad r\to\infty,\label{eq:31}
\end{align}
where
\begin{align}
  |N_{\kappa}|=\sqrt{\frac{\kappa\sinh(\nu\pi)}{\nu}}.\label{eq:32}
\end{align}
Here $\kappa=\sqrt{2m|E_{\text{rel}}|/\hbar^{2}}>0$,
$\nu=\sqrt{\lambda_{\text{c}}-\lambda}>0$, and $K_{i\nu}$ is the
modified Bessel function of the second kind. It is well known
\cite{Case:1950an} that Eq.~\eqref{eq:31} provides orthogonal
functions if $\kappa$ is quantized as
$\kappa_{\ell}=\kappa_{\ast}\e^{-\frac{\ell\pi}{\nu}}$, where
$\kappa_{\ast}>0$ is a newly emerged inverse length scale and
$\ell\in\mathbb{Z}$. Thus, there exist infinitely many negative energy
eigenvalues given by
\begin{align}
  E_{\text{rel}}^{(\ell)}=-\frac{\hbar^{2}\kappa_{\ast}^{2}}{2m}\exp\left(-\frac{2\ell\pi}{\nu}\right),\quad\ell\in\mathbb{Z}.\label{eq:33}
\end{align}
These are the binding energies of $n$-body bound states of identical
particles in the channel $\lambda(<\lambda_{\text{c}})$. Note that
Eq.~\eqref{eq:33} satisfies the geometric scaling
$E_{\text{rel}}^{(\ell+1)}=\e^{-\frac{2\pi}{\nu}}E_{\text{rel}}^{(\ell)}$,
which---as discussed in the introduction---is a manifestation of
discrete scale invariance in the bound-state problem. It should also
be noted that, under the full discrete scale invariance which forms
the group $\mathbb{Z}$, the energy spectrum \eqref{eq:33} cannot be
bounded from below. In other words, in order to make the spectrum
lower-bounded, we have to break this invariance under
$\mathbb{Z}$. One easy way to achieve this is to cut off and
regularize the inverse-square potential. Since the purpose of this
paper is to demonstrate the full discrete scale invariance, we will
not discuss this regularization procedure any further. For more
details, we refer to the literature
\cite{Camblong:2000ec,Camblong:2000ax,Coon:2002sua,Bawin:2003dm,Braaten:2004pg,Braaten:2004rn,Hammer:2005sa,Essin:2006}
(see also \cite{Bedaque:1998kg} for the related field-theory
approach).

\subsection{Exact \texorpdfstring{\boldmath$n$}{n}-body S-matrix
  elements}
\label{section:4B}
Let us next consider the case $E_{\text{rel}}>0$. In this case the
solution to the hyperradial equation \eqref{eq:28c} is given by the
following linear combination:
\begin{align}
  R_{k\lambda}(r)
  &=\sqrt{\frac{\pi kr}{2}}\left(\e^{-i\frac{(1+2i\nu)\pi}{4}}H^{(2)}_{i\nu}(kr)+S_{\lambda}(k)\e^{+i\frac{(1+2i\nu)\pi}{4}}H^{(1)}_{i\nu}(kr)\right)\nonumber\\
  &\to\e^{-ikr}+S_{\lambda}(k)\e^{+ikr}\quad\text{as}\quad r\to\infty,\label{eq:34}
\end{align}
where $k=\sqrt{2mE_{\text{rel}}/\hbar^{2}}>0$. $H^{(1)}_{i\nu}$ and
$H^{(2)}_{i\nu}$ are the Hankel functions of the first and second
kind, respectively. It is straightforward to show that
Eqs.~\eqref{eq:31} and \eqref{eq:34} become orthogonal if the
coefficient $S_{\lambda}(k)$ takes the following
form:\footnote{Similar (but not equivalent) results in terms of the
  phase shift $\delta(k)=\frac{1}{2i}\log S_{\lambda}(k)$ were
  discussed in \cite{Camblong:2000ec,Coon:2002sua,Essin:2006}.}
\begin{align}
  S_{\lambda}(k)=i\frac{\sinh(\nu\pi/2-i\nu\log(k/\kappa_{\ast}))}{\sinh(\nu\pi/2+i\nu\log(k/\kappa_{\ast}))}.\label{eq:35}
\end{align}
This is the S-matrix element of $n$-body scattering of identical
particles in the channel $\lambda(<\lambda_{\text{c}})$. Indeed, it
satisfies the following desired properties:\footnote{Note that
  $S_{\lambda}(k)$ does not fulfill the real analyticity
  property. Instead, it satisfies the interesting identity
  $\overline{S_{\lambda}(k)}=-S_{\lambda}(\kappa_{\ast}^{2}/k)$.}
\begin{itemize}
\item\textit{Unitarity.}
  \begin{align}
    \overline{S_{\lambda}(k)}S_{\lambda}(k)=1,\quad\forall k>0.\label{eq:36}
  \end{align}
\item\textit{Discrete scale invariance.}
  \begin{align}
    S_{\lambda}(\e^{\frac{\pi}{\nu}}k)=S_{\lambda}(k),\quad\forall k>0.\label{eq:37}
  \end{align}
\item\textit{Bound-state poles and residues.}
  \begin{align}
    \lim_{k\to i\kappa_{\ell}}(k-i\kappa_{\ell})S_{\lambda}(k)=i|N_{\kappa_{\ell}}|^{2},\quad\forall\ell\in\mathbb{Z}.\label{eq:38}
  \end{align}
\end{itemize}
Here the overline stands for the complex conjugate. Note that
Eq.~\eqref{eq:37} is equivalent to the log-periodicity of
$S_{\lambda}(k)$ with the period $\pi/\nu$ as a function of $\log
k$. As discussed in the introduction, this log-periodicity is a
manifestation of discrete scale invariance in the scattering problem.

To summarize, we have seen that discrete scale invariance manifests
itself in the geometric series of $n$-body bound states as well as in
the log-periodic oscillation of $n$-body S-matrix elements. Note,
however, that these exact results are based on the assumption that
there exists at least one negative eigenvalue
$\lambda(<\lambda_{\text{c}})$ in the Laplace equation
\eqref{eq:28b}. Let us finally investigate whether and when such a
negative eigenvalue appears. To simplify the problem, below we will
focus on the case $n=3$, in which the critical value is
$\lambda_{\text{c}}=0$.

\section{Example: Exact phase diagram in the three-body problem}
\label{section:5}
Now we wish to solve the Laplace equation on the three-body
hyperangular space
$\Omega_{1}=\{(\Hat{\xi}_{1},\Hat{\xi}_{2}):\Hat{\xi}_{1}^{2}+\Hat{\xi}_{2}^{2}=1,~0<\Hat{\xi}_{1}<\sqrt{3}\Hat{\xi}_{2}\}$
with the scale-invariant boundary conditions.  To this end, let us
first introduce the following polar coordinates in $\Omega_{1}$ (see
Fig.~\ref{figure:1b}):
\begin{align}
  (\Hat{\xi}_{1},\Hat{\xi}_{2})=(\sin\theta,\cos\theta),\label{eq:39}
\end{align}
where $\theta\in(0,\tfrac{\pi}{3})$. Then the Laplace equation
\eqref{eq:28b} simply becomes
\begin{align}
  -\frac{\partial^{2}}{\partial\theta^{2}}\Theta(\theta)=\lambda\Theta(\theta).\label{eq:40}
\end{align}
Note that the codimension-1 boundaries
$\partial\mathcal{M}^{\text{2-body}}_{3,1}$ and
$\partial\mathcal{M}^{\text{2-body}}_{3,2}$ correspond to $\theta=0$
and $\theta=\frac{\pi}{3}$, respectively. The inward-pointing unit
normal vectors on these boundaries are $\bm{n}_{1}=\bm{e}_{\theta=0}$
and $\bm{n}_{2}=-\bm{e}_{\theta=\frac{\pi}{3}}$, where
$\bm{e}_{\theta}$ stands for the unit vector in the $\theta$-direction
at the angle $\theta$. Since the gradient of scalar functions on
$\Omega_{1}$ is
$\bm{\nabla}_{\Omega_{1}}=\bm{e}_{\theta}\frac{\partial}{\partial\theta}$,
the boundary conditions \eqref{eq:25} read
\begin{subequations}
  \begin{align}
    +\frac{\partial}{\partial\theta}\Theta(\theta)-\frac{1}{g_{1}}\Theta(\theta)&=0\quad\text{at}\quad\theta=0,\label{eq:41a}\\
    -\frac{\partial}{\partial\theta}\Theta(\theta)-\frac{1}{g_{2}}\Theta(\theta)&=0\quad\text{at}\quad\theta=\tfrac{\pi}{3},\label{eq:41b}
  \end{align}
\end{subequations}
where $g_{1}$ and $g_{2}$ are real constants. (Note that, for
$n\geq3$, $g_{j}(\Hat{\bm{\xi}})$ in Eq.~\eqref{eq:24} becomes
constant at the boundaries.)

\begin{figure}[t]
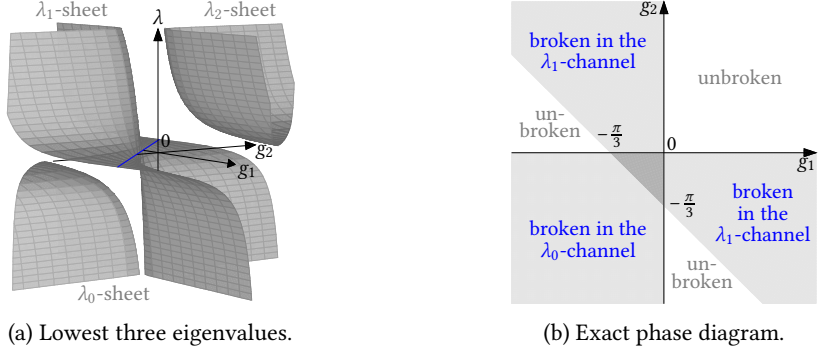

  \centering%
  \begin{subfigure}[t]{0.4\linewidth}
    \centering%
    \input{figure2a.eepic}%
    \caption{Lowest three eigenvalues.}
    \label{figure:2a}
  \end{subfigure}\quad
  \centering%
  \begin{subfigure}[t]{0.4\linewidth}
    \centering%
    \input{figure2b.eepic}%
    \caption{Exact phase diagram.}
    \label{figure:2b}
  \end{subfigure}
  \caption{Parameter dependence of the eigenvalues
    $\lambda_{0}<\lambda_{1}<\lambda_{2}<\cdots$. (a) $\lambda_{2}$ is
    always positive. $\lambda_{1}$ changes sign if we go across the
    blue line defined by $g_{1}+g_{2}=-\frac{\pi}{3}$ on the
    $\lambda=0$ plane. The negative $\lambda_{1}$ exists in the domain
    $D_{1}=\{(g_{1},g_{2}):-\frac{\pi}{3}<g_{1}+g_{2}<|g_{1}-g_{2}|\}$.
    $\lambda_{0}$ is always negative and exists in the domain
    $D_{0}=\{(g_{1},g_{2}):g_{1}<0,~g_{2}<0\}$. (b) The gray-shaded
    region represents the union $D_{0}\cup D_{1}$ in which continuous
    scale invariance is broken to discrete scale invariance in the
    $\lambda_{0}$- or $\lambda_{1}$-channel. The dark-gray region
    represents the intersection $D_{0}\cap D_{1}$ in which continuous
    scale invariance is broken in both the $\lambda_{0}$- and
    $\lambda_{1}$-channels.}
  \label{figure:2}
\end{figure}

Let us now solve the eigenvalue equation \eqref{eq:40} with the above
boundary conditions. First, the general solution for $\lambda\neq0$ is
given by
\begin{align}
  \Theta_{\lambda}(\theta)=A(\lambda)\e^{i\sqrt{\lambda}\theta}+B(\lambda)\e^{i\sqrt{\lambda}(\frac{\pi}{3}-\theta)},\label{eq:42}
\end{align}
where $A(\lambda)$ and $B(\lambda)$ are integration constants. By
imposing the boundary conditions \eqref{eq:41a} and \eqref{eq:41b}, we
get the following quantization condition for $\lambda$
\cite{Fujimoto:2011kf}:
\begin{align}
  \tan\left(\frac{\pi}{3}\sqrt{\lambda}\right)=\frac{(g_{1}+g_{2})\sqrt{\lambda}}{g_{1}g_{2}\lambda-1},\label{eq:43}
\end{align}
or, equivalently,
\begin{align}
  (X-X_{0}(\lambda))^{2}-Y^{2}=Z(\lambda),\label{eq:44}
\end{align}
where $X=\frac{g_{1}+g_{2}}{\sqrt{2}}$,
$Y=\frac{-g_{1}+g_{2}}{\sqrt{2}}$,
$X_{0}(\lambda)=\sqrt{\frac{2}{\lambda}}\cot(\frac{\pi}{3}\sqrt{\lambda})$,
and
$Z(\lambda)=\frac{2}{\lambda}(1+2\cot^{2}(\frac{\pi}{3}\sqrt{\lambda}))$. Note
that Eq.~\eqref{eq:44} defines infinitely many two-dimensional sheets
in the $(g_{1},g_{2},\lambda)$-space whose intersection with the
$\lambda=\text{const}$ plane is a hyperbola; see
Fig.~\ref{figure:2a}. As one can observe from this figure, there exist
two distinct sheets---the $\lambda_{0}$-sheet and the
$\lambda_{1}$-sheet---on which the eigenvalues $\lambda_{0}$ and
$\lambda_{1}$ go below the critical value
$\lambda_{\text{c}}=0$. Close inspection shows that $\lambda_{0}$ is
always negative and exists only in the domain
$D_{0}=\{(g_{1},g_{2}):g_{1}<0,~g_{2}<0\}$; while $\lambda_{1}$
changes sign if it crosses the line $g_{1}+g_{2}=-\tfrac{\pi}{3}$ and
becomes negative in the domain
$D_{1}=\{(g_{1},g_{2}):-\tfrac{\pi}{3}<g_{1}+g_{2}<|g_{1}-g_{2}|\}$. Hence,
in the region $D_{0}\cup D_{1}$, continuous scale invariance is broken
down to discrete scale invariance in the $\lambda_{0}$- or
$\lambda_{1}$-channel. The exact phase diagram is depicted in
Fig.~\ref{figure:2b}. It should be noted that the zero-temperature
transition from the unbroken phase to the broken phase is nothing but
the Berezinskii-Kosterlitz-Thouless-like transition discussed in
\cite{Kaplan:2009kr}.

\begin{figure}[t]
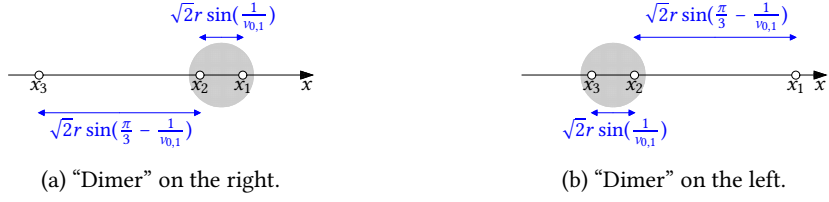

  \centering%
  \begin{subfigure}[t]{0.4\linewidth}
    \centering%
    \input{figure3a.eepic}%
    \caption{``Dimer'' on the right.}
    \label{figure:3a}
  \end{subfigure}\quad
  \centering%
  \begin{subfigure}[t]{0.4\linewidth}
    \centering%
    \input{figure3b.eepic}%
    \caption{``Dimer'' on the left.}
    \label{figure:3b}
  \end{subfigure}
  \caption{Typical particle configurations for (a) the ``dimer'' on
    the right and (b) the ``dimer'' on the left. The sizes of the
    ``dimers'' are about $\sqrt{2}r\sin(1/\nu_{0,1})$, which follow
    from the exponential decay rates $1/\nu_{0,1}$ in
    Eqs.~\eqref{eq:45a} and \eqref{eq:45b} and the relations
    $\frac{x_{1}-x_{2}}{\sqrt{2}}=r\sin\theta$ and
    $\frac{x_{2}-x_{3}}{\sqrt{2}}=r\sin(\frac{\pi}{3}-\theta)$. (Note
    that, in the three-body problem, the relations between the
    normalized Jacobi coordinates and the hyperspherical coordinates
    are $\xi_{1}=\frac{x_{1}-x_{2}}{\sqrt{2}}=r\sin\theta$ and
    $\xi_{2}=\frac{x_{1}+x_{2}-2x_{3}}{\sqrt{6}}=r\cos\theta$, which
    lead to
    $\frac{x_{2}-x_{3}}{\sqrt{2}}=r\sin(\frac{\pi}{3}-\theta)$.)}
  \label{figure:3}
\end{figure}

Before closing this section, it is worthwhile to revisit the
three-body scattering by using the hyperangular wavefunction
\eqref{eq:42}. To simplify the argument, let us consider the case
$g_{1}=g_{2}=:g\in(-\frac{\pi}{6},0)$, in which there appear two
negative eigenvalues $\lambda_{0}$ and $\lambda_{1}$. In this case, it
is easy to show that the eigenfunctions take the following simple
forms:
\begin{subequations}
  \begin{align}
    \Theta_{\lambda_{0}}(\theta)&\propto\e^{-\nu_{0}\theta}+\e^{-\nu_{0}(\frac{\pi}{3}-\theta)},\label{eq:45a}\\
    \Theta_{\lambda_{1}}(\theta)&\propto\e^{-\nu_{1}\theta}-\e^{-\nu_{1}(\frac{\pi}{3}-\theta)},\label{eq:45b}
  \end{align}
\end{subequations}
where $\nu_{0}=\sqrt{|\lambda_{0}|}$ and
$\nu_{1}=\sqrt{|\lambda_{1}|}$ are the solutions to the conditions
$g=-\frac{1}{\nu}\coth(\frac{\pi}{6}\nu)$ and
$g=-\frac{1}{\nu}\tanh(\frac{\pi}{6}\nu)$. Notice that the first terms
in Eqs.~\eqref{eq:45a} and \eqref{eq:45b} sharply localize to the
boundary $\theta=0$ (i.e., the two-body coincidence point
$x_{1}=x_{2}$) with the exponential decay rate $1/\nu_{0,1}$, while
the second terms to the opposite boundary $\theta=\frac{\pi}{3}$
(i.e., $x_{2}=x_{3}$) with the same exponential decay rate. In other
words, these eigenfunctions describe the superpositions of two
``dimers'' whose spatial extents are about
$\sqrt{2}r\sin(1/\nu_{0,1})$; see Fig.~\ref{figure:3}. The physical
meaning of the scattering solutions
$\Theta_{\lambda_{0,1}}(\theta)R_{k\lambda_{0,1}}(r)$ is now clear:
they describe the superpositions of ``atom-dimer'' scatterings. Note,
however, that these scatterings are note quite the same as the
standard atom-dimer scattering in the three-dimensional Efimov effect
\cite{Braaten:2004rn} because the spatial extents of our ``dimers''
scale with the hyperradius $r$; that is, the dimer's size becomes
smaller and smaller as the third particle approaches the
``dimer''. This difference comes from the cluster properties of the
models: in the standard atom-dimer scattering, the total Hamiltonian
decomposes into the summation of the one-body cluster Hamiltonian,
two-body cluster Hamiltonian, and intercluster potential between one-
and two-body clusters, thereby determining the dimer's size only
through the two-body cluster Hamiltonian. In the present model,
however, there is no such cluster decomposition such that the whole
three-body system simply scales with $r$. This scaling is the
characteristic feature of the three-body scattering in our model.

\section{Conclusion}
\label{section:6}
In this paper, we have introduced the most general scale-invariant
model of $n$ identical spinless particles in one dimension, where
interparticle interactions are only two-body contacts. In this model,
we have found that continuous scale invariance can be broken down to
discrete scale invariance for any $n\geq3$. The physical consequences
of this scale-invariance breaking are the onset of geometric series of
$n$-body bound states and the log-periodic oscillation of the $n$-body
S-matrix elements. Thanks to the boson-fermion duality, our findings
can be applied equally well to both bosons and fermions. We emphasize
that our results are based on the assumption that the system fulfills
(i) probability conservation, (ii) permutation invariance, (iii)
translation invariance, and (iv) scale invariance. Hence any $n$-body
problems under two-body contact interactions that satisfy (i)--(iv)
must fall into our model. It should also be emphasized that we did not
require cluster separability in the present paper. If, in addition,
one requires cluster separability, the available parameter space
becomes a much smaller subspace. For example, if we required the
three-body system to be decomposed into the one- and two-body
clusters, then the scale invariance would be realized only for the
Dirichlet and Neumann boundary conditions at the two-body coincidence
points, which correspond to $g_{j}=0$ and $g_{j}=\infty$ in
Eqs.~\eqref{eq:41a} and \eqref{eq:41b}, respectively. Hence in this
case the system cannot exhibit discrete scale invariance, which is
consistent with the no-go result \cite{Nielsen:2001} that the Efimov
effect cannot be realized in one dimension. Our nontrivial results are
therefore applicable to $n$-body systems that cannot be decomposed
into smaller subclusters.

Let us finally comment on the criterion \eqref{eq:30}. Our exact
results \eqref{eq:33} and \eqref{eq:35} in the broken phase are based
on the assumption that there exists at least one negative eigenvalue
$\lambda$ that satisfies the inequality
$\lambda<\lambda_{\text{c}}=-\frac{(n-3)^{2}}{4}$. For $n=3$, we have
checked that this condition is indeed satisfied and determined the
exact phase diagram of scale-invariance breaking. However, the case
$n\geq4$ is left open. From the physical viewpoint, it is quite
reasonable to expect that, for sufficiently strong attractive
interactions, there would always appear at least one negative
eigenvalue $\lambda(<\lambda_{\text{c}})$ for any $n$. This is simply
because, just as in the case of the ordinary $\delta$-function
potential problem in one dimension, at least a single negative
eigenvalue should appear and take an arbitrary large absolute value as
we increase the strength of attractive coupling constants. Future
studies should investigate whether and when the criterion is met for
$n\geq4$ by employing the spectral analysis of the Laplace equation
\eqref{eq:28b} with the scale-invariant boundary conditions
\eqref{eq:25}.

\printbibliography
\end{document}